\documentclass[letter]{aa} 
\usepackage{graphicx}
\usepackage{txfonts}
\usepackage{color}
\definecolor{gray}{rgb}{0.5,0.6,0.7}
\definecolor{raspberry}{rgb}{0.7,0.2,0.4}
\definecolor{darkred}{rgb}{0.75,0.,0.2}

\newcommand{\msun}{\mathrm{M}_\odot}

\begin{document}

\title{The Milky Way accretion history compared to cosmological simulations}

   \subtitle{From bulge formation to dwarf galaxy infall}
   
\titlerunning{MW accretion history versus cosmological simulations} 
\authorrunning{F. Hammer et al.}

   \author{F. Hammer\inst{1},
          Y. J. Jiao\inst{1}, G. A. Mamon\inst{2}, Y. B. Yang\inst{1}, I. Akib\inst{1}, P. Amram\inst{3}, H. F. Wang\inst{4}, J. L. Wang\inst{1}, L. Chemin\inst{5} }

   \institute{GEPI, Observatoire de Paris, PSL University, CNRS, Place Jules Janssen, 92195 Meudon, France
              \email{francois.hammer@obspm.fr}
         \and  Institut d'Astrophysique de Paris (UMR7095: CNRS \& Sorbonne Universit\'e), 98 bis Bd Arago, 75014, Paris, France
          \and Aix-Marseille Univ., CNRS, CNES, LAM, 38 rue Fr\'ed\'ric Joliot Curie, 13338 Marseille, France
          \and Dipartimento di Fisica e Astronomia Galileo Galilei,Universita di Padova,Vicolo Osservatorio 3,I-35122,Padova,Italy
          \and Instituto de Astrofisica, Facultad de Ciencias Exactas, Universidad Andres Bello, Fernandez Concha 700, Las Condes, Santiago RM, Chile
             }

   \date{Received July 5, 2024; }

 
  \abstract
   {Galactic halos are known to grow hierarchically, inside out. This implies a correlation between the infall lookback time of satellites and their binding energy. Cosmological simulations predict a linear relation between the infall lookback time and the logarithm of the binding energy, with a small scatter. {\it Gaia} measurements of the bulk proper motions of globular clusters and dwarf satellites of the Milky Way are sufficiently accurate to establish the kinetic energies of these systems. Assuming the gravitational potential of the Milky Way, we can deduce the binding energies of the dwarf satellites and those of the galaxies that were previously accreted by the Milky Way. This can be compared to cosmological simulations for the first time. The relation of the infall lookback time versus binding energy we found in a cosmological simulation matches that for the early accretion events when the simulated total Milky Way mass within 21 kpc was rescaled to 2 $\times\ 10^{11}\,\msun$. This agrees well with previous estimates from globular cluster kinematics and from the rotation curve. However, the vast majority of the dwarf galaxies are clear outliers to this rescaled relation, unless they are very recent infallers. In other words, the very low binding energies of most dwarf galaxies compared to Sgr and previous accreted galaxies suggests that most of them were accreted much later than 8 or even 5 Gyr ago. We also found that the subhalo systems in some cosmological simulations are too dynamically hot when they are compared to identified Milky Way substructures. This leads to an overestimated impact of satellites on the Galaxy rotation curve.} 


   \maketitle
   
%

\section{Introduction}
In the hierarchical scenario of galaxy formation, structures grow inside out \citep{Gott1975}. This means that newcomers have lower binding energies than satellites that entered a main galaxy host at early epochs. Because of the host mass growth, first-comers are naturally most strongly bound \citep{Rocha2012,Boylan-Kolchin2013}. Cosmological simulations recovered a tight linear correlation between the lookback infall time and the logarithm of the binding energy, showing an evolution of more than 1 dex, and with a scatter of only 0.13 dex \citep{Rocha2012}. The slope of this correlation is consistent with the slope estimated for the Milky Way (MW) accretion history. Assuming the MW mass model from \citet[total mass of $8.3 \times 10^{11}\,\msun$]{Eilers2019}, \citet[see their Fig. 6]{Hammer2023} determined this relation for the MW. The authors accounted for globular clusters (GCs) associated with the bulge, Kraken, Gaia-Sausage-Enceladus (GSE), and Sgr infall and adopted the associations made by \citet{Malhan2022} and \citet{Kruijssen2020}. According to the latter study, these events occurred 12.5$\pm$0.5, 11.5$\pm$0.5, 9$\pm$1, and 5$\pm$1 Gyr ago, respectively. The relation between their lookback infall time and the logarithm of their binding energy was found to be linear, which agrees with cosmological simulations. 

The relation was used by \citet{Hammer2023} to infer that the binding energies of MW dwarf galaxies are far lower than the binding energy of GSE GCs (by a factor of 6 on average). This prevented them from having been accreted $\sim$ 9 Gyr ago. Extrapolating the relation from bulge to Sgr GCs toward the dwarf energy regime, \citet{Hammer2023} concluded that most dwarf infall lookback times should be shorter than 3 Gyr. The relation could also constrain the MW mass, because the more massive a galaxy, the deeper its potential well. This allows it to capture satellites with lower binding energies (e.g., Leo I could be bound if the MW were very massive; \citealt{Boylan-Kolchin2013}). Recent {\it Gaia} measurements of the MW rotation curve (RC) provided $8.3 \times 10^{11}\,\msun$ \citep{Eilers2019} and $2.06^{+0.24}_{-0.13}\times 10^{11}\,\msun$ \citep{Jiao2023} for an adopted \citet[Navarro, Frenk \& White, hereafter NFW]{Navarro1997} and \citet[see also \citealt{Retana-Montenegro2012}]{Einasto1965} mass profile, respectively. However, the 
 NFW mass profile is too shallow in the disk outskirts, and it is rejected at 3$\sigma$ level  by  RC measurements from the 3rd {\it Gaia} data release (hereafter {\it Gaia} DR3), which also revealed a decline beyond 19 kpc that is consistent with Keplerian expectations \citep{Wang2023,Jiao2023,Ou2024}. The consideration of the MW gravitational potential as being almost equivalent to a 
point mass\footnote{The best-fit Einasto index found by \cite{Jiao2023} is consistent with a Gaussian cutoff of the dark matter density profile.} beyond 19 kpc has profoundly impacted the cosmological community, for whom the MW halo-limiting radius was generally assumed to range from 150 to 300 kpc.

However, the MW is not entirely axisymmetric or in dynamical equilibrium, as is assumed when resolving the Jeans equation to derive the rotation curve. In particular, the disk shows many substructures, including ridges, warps, and flares, in particular, in its outer range, with different upward or inward velocities, and a significant difference in the stellar rotational velocity above and below the disk \citep[and references therein]{Antoja2021}. Recently, \citet{Koop2024} suggested that the MW disk is so perturbed by passages of satellites that its rotation curve cannot be used to predict its mass. However, the regularity of the MW rotation curve until $\sim$19 kpc suggests that nonequilibrium mechanisms are not sufficiently strong to perturb its dynamics significantly, although they could be a serious concern beyond this radius.

The goal of this paper is to verify which MW mass can be consistent with the relation between lookback infall time and the binding energy of satellites after we compare the mass to the masses obtained in cosmological simulations. Furthermore, we assess the accretion epoch of dwarf galaxies, and we verify whether satellites may impact the disk stability when they pass near the disk.\\
Section~\ref{cosmo_simu} tests the accretion history of the MW, including that of dwarf galaxies, and compares it to the history determined from high-resolution cosmological simulations. Section~\ref{discussion} discusses the pertinence of cosmological simulations for a retrieval of the MW properties and its accretion history, and for identifying whether satellites may have affected its disk dynamics. We then summarize our main conclusions.

In this paper, $R_{\rm 200c}$ is the virial radius for which the enclosed DM mass, dubbed $M_{\rm 200c}$, corresponds to an overdensity of 200 times the critical density. The total mass includes $M_{\rm 200c}$, to which we added the baryonic component, in a similar way as \citet{Eilers2019}.  The orbital energies correspond to the addition of the kinetic energy provided by {\it Gaia} proper motions and radial velocities with the potential energy of the considered mass distribution. To calculate the potential, we followed \citet{Rocha2012}, who assumed that the DM-halo potential rises to zero at 1 Mpc. We therefore assumed that the DM potential associated with a given virial mass $M_{\rm 200c}$ must reach zero at the rescaled radius of $1\,(M_{\rm 200c}/1.4 \times 10^{12})^{1/3} \,\rm Mpc$, where 1.4 $\times 10^{12}\,\msun$ is the DM-halo mass $M_{\rm 200c}$ of \citeauthor{Rocha2012}\footnote{The $M_{\rm 200c}$ value was converted from the \citet{Rocha2012} $M_{\rm 200m} =1.9 \times 10^{12}\,\msun$, which corresponds to an overdensity of 200 times the mass density.}. The total mass (or potential) is the sum over the DM and the bulge and disk components. As an example, detailed equations are given in Appendix~\ref{app:rescaling} for an NFW DM-halo component.

   \begin{figure}
   \centering
 \includegraphics[width=8cm]{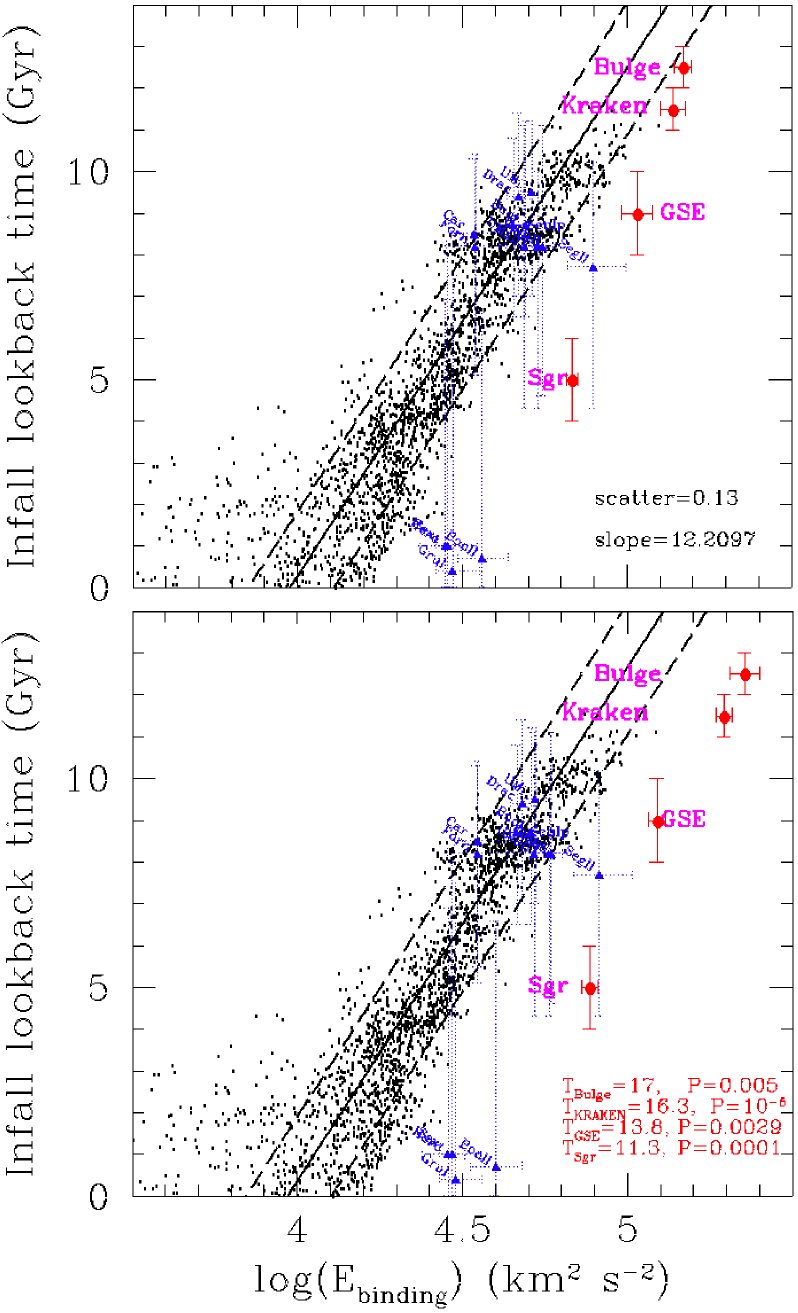}
   \caption{Relation between the infall lookback time and the logarithm of the binding energy, for which black dots represent the subhalos that are associated with the \citet{Rocha2012} simulation, the solid line represents the best fit, and the dashed lines show its one $\sigma$ scatter. The binding energies of dwarf galaxies (blue triangles) and of GCs associated with MW events (red dots with corresponding labels in magenta) are calculated using the observed kinetic energy together with the potential of the main halo. The blue triangles indicate the location of dwarf galaxies with a very good accuracy ($<$0.1 dex) in binding energy, and for which we adopted the predicted dwarf infall time of \citet{Barmentloo2023}. Top panel: Relation as derived from the \citet{Rocha2012} dark-matter-only simulations. 
    Bottom panel: Same as the top panel, but we added a baryonic mass (6.2$\times\ 10^{10}\,\msun$) after a slight rescaling of the dark matter halo mass from 1.4 $\times 10^{12}\,\msun$ to 1.34 $\times 10^{12}\,\msun$ (see Appendix~\ref{app:rescaling}). In the bottom right corner of the panel, the numbers indicate the predicted time for GCs associated with the bulge, GSE, and Sgr when they are inserted into the halo potential, as well as the probability that they lie in the distribution of simulated subhalos.
   }
              \label{fig:Rocha}
    \end{figure}

 %

\section{Cosmological simulations compared to the MW RC and accretion history}
\label{cosmo_simu}

\begin{figure}
   \centering
\includegraphics[width=9cm]{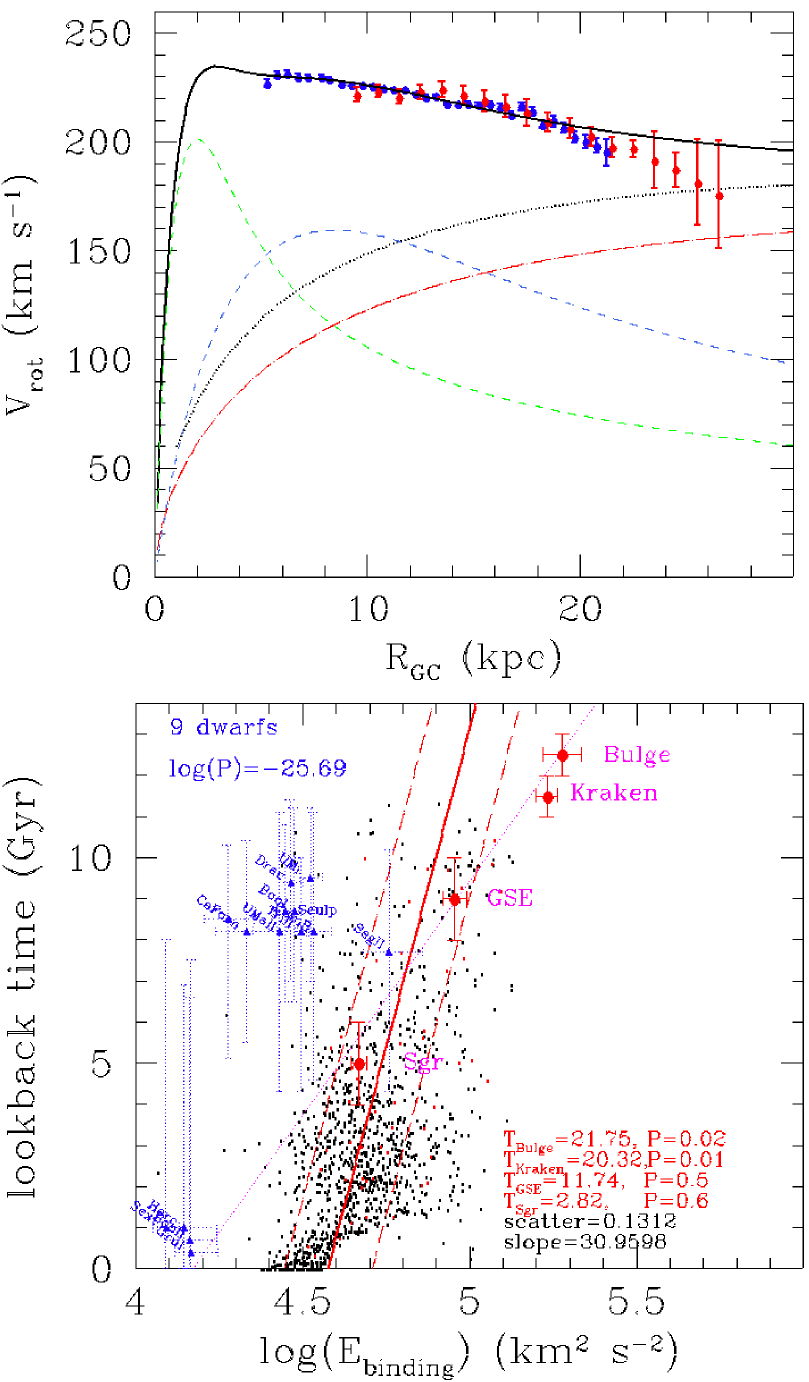}
      \caption{ Rotation curve and MW accretion history for the {\it FIRE} model m12c. Top panel: Rotation curve from an isolated MW-like galaxy from {\it FIRE} (halo m12c) compared to {\it Gaia} data. The blue points show DR2 \citep{Eilers2019} without systematics, and the red points show DR3 \citep{Jiao2023}, which includes measurements from \citet{Wang2023} plus systematic errors. The contributions from the DM halo, bulge, and disk are plotted as long-dash red, short-dash green, and short-dash blue lines, respectively. To illustrate the impact of the very massive halo from \citet{Rocha2012}, we also added its contribution (dotted black line), which far exceeds the {\it FIRE} m12c DM halo. Bottom panel: Same as the bottom panel of Figure~\ref{fig:Rocha}, but for the {\it FIRE} halo m12c. The binding energies of dwarf galaxies and of GCs associated with MW events were calculated using the observed kinetic energy together with the potential of each main halo. The top left corner of the panel displays the combined probability that the nine dwarfs have infall lookback times longer then 8 Gyr, as predicted by \citet{Barmentloo2023}. In the bottom right corner of the panel, the numbers indicate the predicted time for GCs associated with the bulge, GSE, and Sgr when they are inserted into the m12c halo potential, as well as the probability that they lie in the distribution of subhalos. 
              }
         \label{fig:m12c}
   \end{figure}

The top panel of Figure~\ref{fig:Rocha} compares the binding energies of the simulated subhalos according to \citet{Rocha2012} \footnote{The infall time versus binding energy relation of \cite{Rocha2012} is based on a high-resolution ($4100\,\msun$ for dark matter particles) dark-matter-only simulation based on {\it Via Lactea II} \citep{Diemand2007}} (black dots) with those of GCs that are associated with the bulge, GSE, and Sgr (red dots). The \citet{Rocha2012} model has a high total mass ($M_{\rm 200c}$ = 1.4 $\times 10^{12}\,\msun$). However, this model faces two fundamental difficulties: (1) The binding energies of the simulated subhalos are lower by two to three times than those of Sgr and the GSE GCs, and to reconcile them, extremely high lookback time values are required for these MW substructures. (2) The dark matter (DM) halo is far too massive to fit the MW RC (see the top panel of Figure~\ref{fig:m12c} and compare the dotted black line to the long-dash red line). The addition of a baryonic component to the simulations of \citet{Rocha2012} would not help us to reproduce  the locations of bulge, Kraken, GSE, and Sgr points (see the bottom panel of Figure~\ref{fig:Rocha}) because a concentrated mass component would worsen the offset by increasing their binding energy.

In Appendix~\ref{app:rescaling} we tried to rescale the simulations by \citet{Rocha2012} to different values of the total mass within the virial radius. We failed to find a value for which they fit the locations of GCs associated with Sgr, GSE, Kraken, and the bulge, however. Thus, while \citet{Rocha2012} reproduced the slope of the relation of infall time versus binding energy well, the kinetic energy of the simulated subhalos is far higher than that of typical GC systems associated with Sgr, GSE, Kraken, and the bulge even after the total mass is rescaled.

%
   %

It is then useful to compare the MW accretion history to high-resolution simulated halos with a full treatment of the baryonic physics. We examined and extracted infall times and binding energies for satellites around Milky Way-mass galaxies in different high-resolution zoom-in hydrodynamical cosmological simulations: Seven from  {\it FIRE} \citep{Wetzel2023}, and 15 from {\it Auriga} \citep{Grand2024} simulated MW-mass galaxies (see details in Appendix~\ref{app:cosmo}). Most {\it FIRE}-simulated MW-mass galaxies do not reproduce the RC slope at large radii (see Figure~\ref{fig:m12c} and Figure~\ref{FigB3}), which is expected because their halo radii are large ($R_{200c}$ from 168 to 251 kpc), which implies significant amounts of mass outside of $\sim$ 20 kpc. They do not reproduce the slope of the relation between infall epoch and binding energy either, which ranges from 27 to 37 Gyr $\rm km^{-2}s^2$, that is, two to three times larger than observed (12.2 Gyr $\rm km^{-2}s^2$). This motivated us to also investigate {\it Auriga}-simulated MW-mass galaxies, including their low MW-mass subsample. Their low MW mass halos 3 and 8 show a rotation velocity that is consistent with observed values around 15 to 20 kpc (see Figure~\ref{FigB5}). However, their RCs are too flat and cannot provide a reasonable fit of the observed MW RC. Furthermore, even low MW-mass halos (e.g., numberes 1 and 9) underestimate the MW RC velocities, which would require a more concentrated DM component.  Most of the {\it Auriga}-simulated MW-mass galaxies show very to extremely steep slopes for the relation between infall epoch and binding energy and predict binding energies that are $\sim$ 0.4 to 0.8 dex too low at the epochs of the bulge formation and of the Kraken, GSE, and Sgr merging events. This is far more problematic. 

Among all the 22 simulated halos, {\it FIRE} m12c \citep{Garrison-Kimmel2019} was found to reproduce both the MW RC and its accretion history best (see Figure~\ref{fig:m12c}). Its total mass is $M_{\rm 200c}$= 9.12 $\times 10^{11}\,\msun$, which is 10\% higher than in the models adopted by \citet{Eilers2019} or by \citet{Bovy2015}, with which it shares many similarities. The top panel of Figure~\ref{fig:m12c} suggests that this model reproduces the {\it Gaia} DR3 MW RC relatively well (associated $\chi^2$ probability $P=0.33$). When this is compared to the RC from \citet{Eilers2019}, the associated $\chi^2$ probability becomes extremely low because the latter authors did not account for systematic errors. A better estimate requires a significant improvement of the systematic error estimates, which is the subject of a future paper (Jiao et al. in preparation). 
Furthermore, the m12c predicted slope of the relation between the infall epoch and binding energy (31 Gyr $\rm km^{-2}s^2$) is more than twice what is observed in the MW. When we account for the fitting of the MW accretion history, the m12c halo might be discarded with a probability of $P= 2 \times 10^{-4}$, which is much better than any other tested simulated halo, however. We also note that {\it FIRE} halos m12b and m12m might be rescaled in total mass and would then reproduce the relation of infall time versus energy with the same significance as halo m12c, but their slopes would again be too high in this relation.\\

\citet{Barmentloo2023} pioneered the use of the accretion history relation to infer the dwarf galaxy infall times. They used 1628 simulated MW halos from {\it EAGLE} \citep{Grand2017} with total masses from 0.5 to 2 $ \times 10^{12}\,\msun$. We adopted their predicted infall lookback times for 14 dSphs, which include Carina, Draco, Fornax, Sculptor, Sextans, and Ursa Minor (in addition to Bootes I and II, Coma Berenices,  Grus I, Hercules, Segue II, Triangulum II, and Ursa Major II; see the blue triangles in Figure~\ref{fig:Rocha}), for which the binding energy is accurately determined with an error smaller than 0.1 dex. Figure~\ref{fig:Rocha} confirms the prediction by \citet{Rocha2012} that most dSph infall lookback times are within 8 to 9 Gyr. This requires a very high MW mass that is similar to the mass derived when dwarf orbits are assumed to be at equilibrium with the MW potential (e.g., $M_{200c}$ =1.3 $\times 10^{12}\,\msun$ from \citealt{Li2017}). However, a fit of the MW accretion history relation requires a lower mass, such as that of {\it FIRE}  m12c. The bottom panel of  Figure~\ref{fig:m12c} shows that an infall lookback time of 5 to 9 Gyr would lead for most dwarfs to an extremely low probability (see the value in the top left corner), while values below 3 Gyr are very likely, except for Segue II. In other words, when a cosmological simulated halo is able to fit the MW relation between the binding energy and the infall time, it automatically implies very low probabilities for most dwarfs to have entered the halo at a similar epoch as the GSE. The infall times of \citet{Barmentloo2023} have very large uncertainties, and their prediction of dwarf infall times is affected by the use of many halos that do not fit the MW accretion history and RC and by their use of several additional parameters that are insensitive to the infall time (see the discussion in the Appendix of \citealt{Hammer2024}). 

\section{Discussion and summary}
\label{discussion}
We have presented the relation between infall epoch and binding energy, which is representative of the MW accretion history. We discuss whether this relation can be sufficiently robust to be predictive. For example, \citet{Pagnini2023} have questioned the reliability of associating GCs with past merger events in the MW \citep{Kruijssen2019,Kruijssen2020,Malhan2022} after they simulated mergers of two galaxies which had their own GC systems. They found GCs that progressively lost their energy after they were stripped in a 1:10 minor encounter, which prevents their use for a dating of the epoch of their simulated mergers. However, most GCs are expected to be formed during strong star formation events that occurred 12 to 9 Gyr ago \citep{Haywood2016}, that is, during the Kraken and GSE merger events \citep{DeLucia2024,Valenzuela2024}. It is doubtful whether the 1:10 mass ratio adopted by \citet{Pagnini2023} is representative of the two latter events. For example, \citet{Naidu2021} considered higher mass ratios\footnote{Notice that these larger mass ratios are necessary to explain the origin of both thin and thick disk of a spiral galaxy like the MW \citep{Hammer2009,Hammer2018b,Hopkins2010,Athanassoula2016,Sauvaget2018}.} of 1:2 to 1:4. Only a modest fraction of MW GCs can then be accreted through the mechanism proposed by \citet{Pagnini2023}, which might provide a complementary channel for GCs that are not identified inside a structure in the plane of total energy - angular momentum \citep[see their Fig. 5]{Hammer2023}. 
In addition, \citet{DSouza2022} argued that the relation of the infall epoch and binding energy might become very uncertain for halos with very active merger histories. The authors analyzed the {\it ELVIS} suite of 48 simulated host halos with a total mass ranging from 1 to 3 $\times\ 10^{12}\,\msun$ \citep{Garrison-Kimmel2014}. Among them, \citet[see their Fig. 9]{DSouza2022} identified 3 host halos with a rich merger history such that the scatter of the relation almost equaled its amplitude. However, the MW is known for its relatively quiescent merger history when compared to spiral galaxies of similar masses \citep{Hammer2007}. For example, while an average spiral galaxy experimented its last major merger 6 Gyr ago, this occurred 9 Gyr ago (GSE) for the MW. The latter is thus unlikely to be one of the few galaxies with the richest merger history, which agrees with the well-identified linear relation \citep{Hammer2023}.

A comparison of the MW accretion history relation to that of cosmological simulations may provide us an additional constraint on the MW mass. However, the mass constraint only applies on the farthest considered GCs, that is, those attached to Sgr, with an average distance of $\sim$ 21 kpc. Within the latter radius, we find a mass of $2.05 \times 10^{11}\,\msun$  for the m12c simulated halo (see Figure~\ref{fig:m12c}). This value agrees well with estimates for GCs by \citet[$2.05 \times 10^{11}\,\msun$]{Watkins2019} and with the values extracted from Gaia DR2 \citet[$2.15 \times 10^{11}\,\msun$]{Eilers2019} and DR3 \citet[$1.95 \times 10^{11}\,\msun$]{Jiao2023} RCs.\\

 Cosmological simulations that can reproduce the binding energy of GSE GCs (see Figure~\ref{fig:m12c} and the three first panels of Figure~\ref{FigB3}) inferred that dwarf galaxies cannot have been bound to the MW 8 to 10 Gyr ago. It confirms the conjecture by \citet{Hammer2023}  that they are newcomers (infall  lookback time $< $3 Gyr). This also suggests that they are likely out-of-equilibrium systems (see the detailed explanations and modeling in \citealt{Hammer2024} and in \citealt{Wang2024}). This suggests that a derivation of the mass from dwarf galaxy orbits after considering them as long-term MW satellites systematically overestimates the total MW mass. 
This is confirmed by the fact that an MW mass in excess of $10^{12}\,\msun$ can neither reproduce the binding energy of the bulge, GSE, and Sgr GCs (see Figure~\ref{fig:Rocha} and Figure~\ref{FigB3}) nor its RC. 


We compared our results to cosmological hydrodynamical simulations from {\it FIRE} and {\it Auriga} and found the results at odds with both the MW RC and its accretion history relation. Many simulated galaxies predicted an RC and accretion history that are very different from the observed events. In particular, \citet{Rocha2012} and {\it Auriga}-simulated galaxies show dynamically hot subhalo systems with binding energies that are lower by three to nine times than those of the bulge, GSE, and Sgr GCs. This suggests that cosmological simulations would benefit from investigating different initial conditions. For example, it could be useful to investigate different initial mass fluctuations especially toward the low-mass end to allow the possibility of more recent major mergers, as are expected for understanding the formation of spiral galaxies and the acquisition of their angular momentum \citep{Hammer2009,Hopkins2010,Puech2012}.

 Our comparisons of the relation of infall epoch and binding energy are limited to cosmological simulations that might provide a too shallow halo-mass distribution, that is, those following an NFW mass profile. If the {\it Gaia} DR3 RC with a Keplerian decline \citep{Jiao2023,Ou2024}  is confirmed, simulations of more concentrated DM halos with lower total masses would be required, and we would need to verify whether they can reproduce the MW RC and the accretion history. However, it would also be necessary to generate simulated halos that are sharply truncated, as suggested by the MW RC.

Finally, the relation of the infall time to energy may help us to verify whether the MW RC can be significantly altered by the passages of heavy satellites through the disk, as suggested by \citet{Koop2024}\footnote{\citet{Koop2024} also suggested a truncation of the MW disk profile from their Jeans equation analysis of {\it Gaia} data with radial velocities. Their analysis was based on Bayesian techniques for estimating distances, but their statistics beyond R= 17 kpc were lower than in \citet{Ou2024}. We defer this to a discussion in a paper (Jiao et al., in preparation) that further analyzes the systematics associated with the {\it Gaia} DR3 data.}. Their study  compared such an impact after assuming the model of \citet{Laporte2018} of Sgr with an initial mass of 6 $\times\ 10^{10}\,\msun$. However, a much lower initial mass for Sgr (3.8 $\times\ 10^{9}\,\msun$) is required\footnote{Both studies considered a similar MW total mass.} to reproduce the Sgr stream \citep{Vasiliev2021}, which according to the latter authors would need a reanalysis of the role of Sgr in seeding the spiral pattern in phase space that was discovered in the MW disk \citep{Antoja2021}. Concerning the impact of Sgr on the MW RC, we may consider that decreasing a perturbing mass by a factor $\sim$ 16 would reduce the predicted changes in the rotation velocity by \citet{Koop2024} by about four times when the perturbations are in the linear regime. This implies that the changes in rotation velocity of 10 to 15 $\rm km\,s^{-1}$ reported by \citet[see their Fig. 11]{Koop2024} would become consistent with the analyses of the systematic uncertainties ($\sim$ 2\% of the rotation velocity below $R= 22\,\rm kpc$) made by \citet{Jiao2023}. \citet{Koop2024} also considered the six halos from {\it Auriga} (see Figure~\ref{FigB4}) to  also infer large systematics caused by the passage of satellites similar to Sgr. However, {\it Auriga} subhalos are four to six times dynamically hotter than Sgr for a common infall lookback time of 5 Gyr, or in other words, the velocity dispersion of the subhalo system in {\it Auriga} is greater than the typical velocity of Sgr relative to the MW. The impact of the satellites on the MW disk would thus become quite modest and consistent with the expected systematic uncertainties \citep{Sylos_Labini2023,Jiao2023,Ou2024}. A significant advantage of the above analysis is that it naturally explains why all {\it Gaia}  RCs show a smooth decline from $R=8$ to 18 kpc \citep{Eilers2019,Mroz2019,Jiao2023,Ou2024}, that is, why they do not show strong local variations in velocity that would be caused by massive satellite impacts on the MW disk.

\begin{acknowledgements}
We are very grateful for the discussion we have had with Robert Grand. We warmly thank the referee whose comments have helped us to improve the paper. We are grateful for the support of the International Research Program Tianguan, which is an agreement between the CNRS in France, NAOC, IHEP, and the Yunnan Univ. in China. Y.-J.J.  acknowledges financial support from the China Scholarship Council (CSC), No.202108070090. I.~A. would like to thank the Graduate Program in Astrophysics of the Paris Sciences et Lettres (PSL) University for funding.
\end{acknowledgements}

%
\bibliographystyle{aa} 
 \bibliography{main.bib} 
%

\begin{appendix} 

\section{Re-scaling binding energy and orbital time
s for halos of different mass}
\label{app:rescaling}
\begin{figure}
   \centering
\includegraphics[width=8cm]{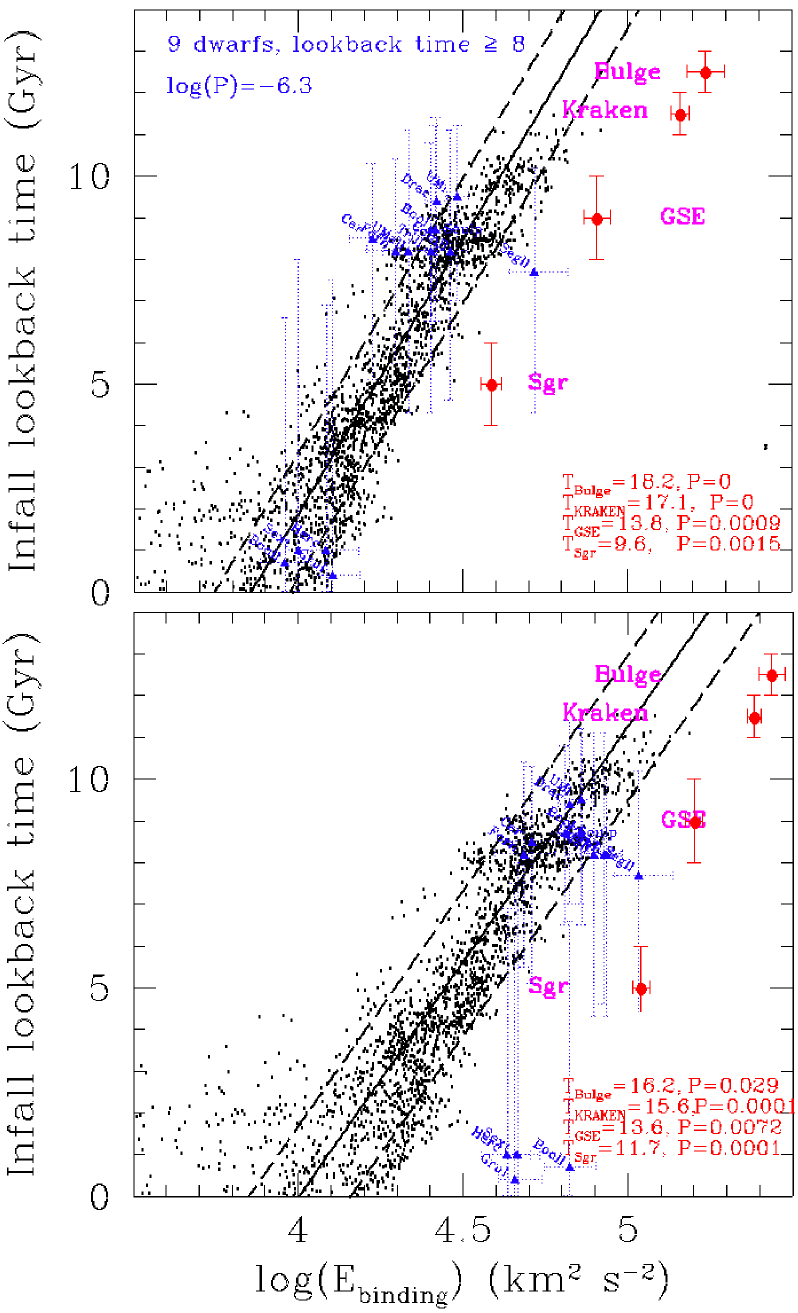}
      \caption{ Same as Fig.~\ref{fig:Rocha}, but for different mass re-scalings of the \citet{Rocha2012} simulations. In the bottom right of each panel the numbers indicate the predicted time for GCs associated with the bulge, GSE, and Sgr when they are inserted into the m12c halo potential, as well as the probability they can lie in the distribution of subhalos. (Top:) We have scaled down the \citet{Rocha2012} main halo mass to 7.4$\times\ 10^{11}\,\msun$ and add a baryonic component of 6.2$\times\ 10^{10}\,\msun$, similar to that used in \citet{Jiao2023} and \citet{Ou2024}.  (Bottom panel:) We have scaled up the \citet{Rocha2012} main halo mass to 19.4$\times\ 10^{11}\,\msun$, also adding a baryonic mass of 6.2$\times\ 10^{10}\,\msun$.
              }
         \label{FigA1}
   \end{figure}

 Figure~\ref{FigA1} illustrates the major problem for the simulations made by \citet{Rocha2012} if it is used to reproduce the MW accretion history after comparing the location of bulge, GSE and Sgr GCs to that of the simulated subhalos. We have added a baryonic component (bulge and disk) very similar to that used by \citet{Jiao2023} and \citet{Ou2024}, as done in the bottom panel of Figure~\ref{fig:Rocha}. The two panels show that for any re-scaling up or down of the \citet{Rocha2012} main halo, 
the subhalos are less bound to their host galaxy (rescaled to a possible Milky Way mass) than the bulge, Kraken, GSE or Sgr GCs are to the Milky Way. Since we kept the kinetic terms of the binding energies of these four systems from the bulk line-of-sight velocities and Gaia bulk proper motions, while adopting the re-scaled potential of the simulation for their potential terms, any rescaling of the simulation will push the positions of these four systems (red points) in the same direction as are pushed the subhalos (black points), although by a smaller amount.  

Here we show that the binding energies and infall times of subhalos in the simulations of \citet{Rocha2012} can be converted for halos having similar concentrations but different masses. Indeed, the binding energy at the `virial' radius $R_{\rm 200c}$ follows $E_{\rm binding}(R_{\rm 200c}) \propto V_{\rm 200c}^2$, where $V_{\rm 200c}^2 = G\,M_{\rm 200c}/R_{\rm 200c}$ is the square of the circular velocity at the virial radius. Since, at any epoch, the mean density within the virial radius is the same for all halos, it leads to $R_{\rm 200c} \propto M_{\rm 200c}^{1/3}$, and $V_{\rm 200c} \propto R_{\rm 200c} \propto M_{\rm 200c}^{1/3}$. One then finds that $E_{\rm binding}(R_{\rm 200c}) \propto M_{\rm 200c}^{2/3}$. 
Since the concentration of halos varies weakly with halo mass ($c \propto M^{-0.1}$, \citealt{Navarro1997}), decreasing the mass by a multiplicative factor  $\mu = 1.9$ leads to only a 6 per cent increase of the concentration. Thus, to first order, re-scaling the mass is self-similar and should not affect the orbital parameters of satellites, if considered in virial units. Therefore, decreasing the mass by a factor of $\mu$ causes their binding energies to be divided  by $\bf \mu^{2/3}$. On the other hand, the circular orbital times of dwarfs around the MW at its virial radius, $t_{\rm orb,circ} \propto R_{\rm 200c}/V_{\rm 200}c$, will be independent of $\mu$. Therefore,  the orbital times for general orbits within the MW halo will be independent of $\mu$, and so will be the infall times, as well as the lookback times from entry to a given fraction of the current virial radius.

In summary, re-scaling the virial mass of the MW by a factor $\mu$ should shift the location of the infall lookback time vs. binding energy of dSphs by a linear factor of $(2/3)\,\log(\mu)$ along the logarithmic energy axis, without affecting the infall lookback time. This will conserve the slope of the correlation. Since \citet{Rocha2012} assumed a potential that rises to zero at 1 Mpc, we have scaled all the considered halo potentials in this paper to reach zero at a radius $R_{\rm lim}$ equal to 1 Mpc multiplied by ($M_{\rm 200c}/1.4 \times 10^{12})^{1/3}$. In Appendix~\ref{app:cosmo} we verify whether models having a full treatment of the baryonic physics may help to identify simulated galaxies having properties similar to that of the MW. \\

Here, we give an example of how the virial mass ($M_{\rm 200c}$) and the potential are calculated for the \citet{Rocha2012} main halo mass of $M_{\rm 200c}$= 14$\times$$10^{11}\,\msun$ that is calculated within the virial radius $R_{200c}$= 231 kpc. 

The DM (spherical) mass profile for our assumed NFW model is:
\begin{equation}
M(R)= M_{200c}\,
\frac{\ln(R/a+1)-R/a\,/\,(R/a+1)}
{\ln(c+1)-c/(c+1)}
\ ,
\end{equation}
where $c=11.48 = R_{200c}/a$ is the concentration, while $a=20.12$ kpc is the NFW scale radius. The gravitational potential corresponding to the NFW model is given by:

\begin{flalign}
\label{eq:potential}
\Phi(R) & = -
\frac{G\,M_{200c}/R_{200c}}{\ln(c+1)-c/(c+1)}
 \nonumber \\
& \qquad 
\times \left[
\frac{\ln(R/a+1)}{R/R_{200c}}
-
\left(\frac{R_{200c}}{R_{\rm lim}}\, \right)
\ln\left(\frac{R_{\rm lim}}{a}+1\right)
\right] \nonumber \\
& = -\frac{G\,M_{200c}/R}{\ln(c+1)-c/(c+1)} \nonumber \\
& \qquad 
\times \left[
\ln\left(\frac{R}{a}+1\right)
-
\left(\frac{R}{R_{\rm lim}}\, \right)
\ln\left(\frac{R_{\rm lim}}{a}+1\right)
\right] 
\ .
\end{flalign}
The last term in the square brackets accounts for the potential going to zero at $R_{\rm lim}$= 1 Mpc. 
The total mass and total potential can then be calculated by adding baryonic mass and potential such as described in \citet{Jiao2023} and \citet{Ou2024}.

\section{Cosmological models compared to the observed MW RC and accretion history}
\label{app:cosmo}

Cosmological hydrodynamical simulations bring widely-used resources providing important predictions for galaxy formation and evolution. In particular "zoom-in" simulations allow to investigate numerical resolutions associated with $10^{4}-10^{5}$ ($10^{3}-10^{4}$) $\msun$ per DM (baryonic) particle \citep{Wetzel2023,Grand2024}, respectively. These simulations include many ingredients of baryonic physics allowing comparison to very detailed observations of the MW substructures and its halo satellites. {\it FIRE} \citep{Wetzel2023} and {\it Auriga} \citep{Grand2024} have considered MW-like (or MW-mass) galaxies with total mass ranging from $M_{\rm 200c}$= 7.3 to 14$ \times 10^{11}\,\msun$, and from 5.3 to 19 $\times 10^{11}\,\msun$, respectively. 
The above mass ranges, while consistent with the value found by {\it Gaia} DR2 RC \citep{Eilers2019}, are  larger than that derived from the {\it Gaia} DR3 RC \citep{Jiao2023,Sylos_Labini2023,Ou2024}. We have tested their predictions on the observed RC and accretion history of the MW, to verify whether this could bring additional constraints on the MW halo mass.We have selected halos from data release of both the {\it FIRE} \citep{Wetzel2023} and {\it Auriga} \citep{Grand2024}.  We have preselected the 7 isolated halos from {\it FIRE}, and 15 {\it Auriga} halos, including 6 high-mass MW with high-resolution, plus the 9 low-mass MW halos.  Our goal was to account for isolated simulated galaxies as well as to account for the simulations with the highest resolution as well as considering the largest possible range for the simulated halo masses.

Each subhalo is represented by its rotation curve (solid black lines in the top panels) that is compared with {\it Gaia} DR3 RCs, and its accretion history relation (bottom panels). In the latter, infall lookback times have been retrieved from the data base for {\it FIRE} simulations, while we have derived them for {\it Auriga} simulations.

\begin{figure}
    \centering
    \includegraphics[width=8cm]{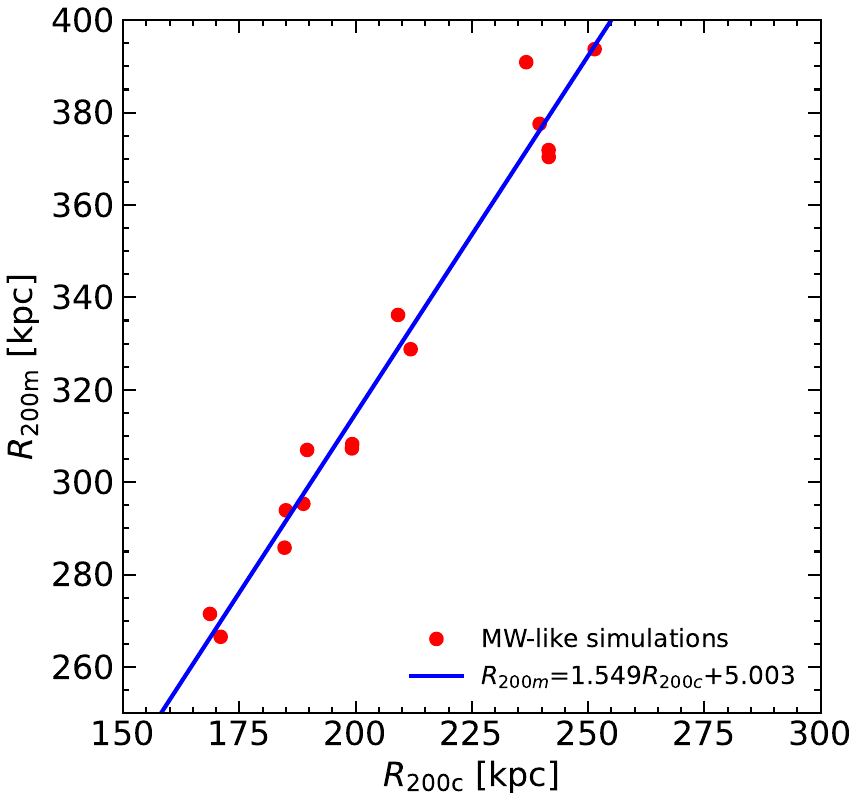}
    \caption{Linear relation of $R_{200m}$ and $R_{200c}$ for different MW-like galaxies realized from {\it Auriga} simulations.
    }
    \label{r200c_r200m}
\end{figure}

\begin{figure}
    \centering
    \includegraphics[width=8cm]{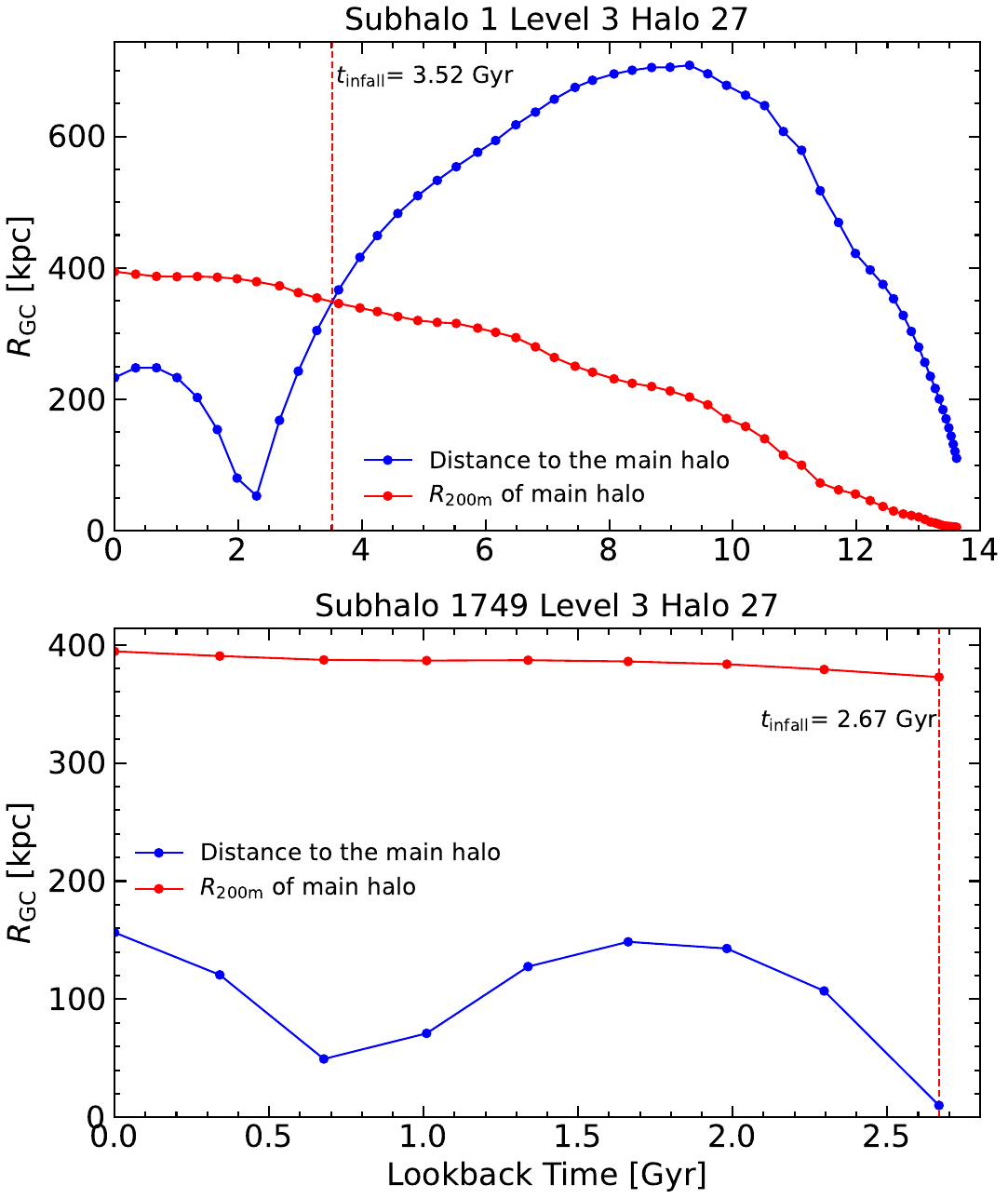}
    \caption{Example of calculation of infall time for {\it Auriga} simulations. Blue points and dash lines indicate the distance of subhalos 1 (top panel) and 1749 (bottom) from {\it Auriga} MW-like Level 3 resolution Halo 27 to the main host galaxy center. Red points and lines indicate the evolution of the virial radius ($R_{200m}$) of the main host galaxy at different epochs using the linear relation in Fig~\ref{r200c_r200m}. Vertical red dashed lines indicate the last infall time of the subhalo.
        }
    \label{r200_lookback}
\end{figure}

\begin{figure*}
   \centering
\includegraphics[width=16cm]{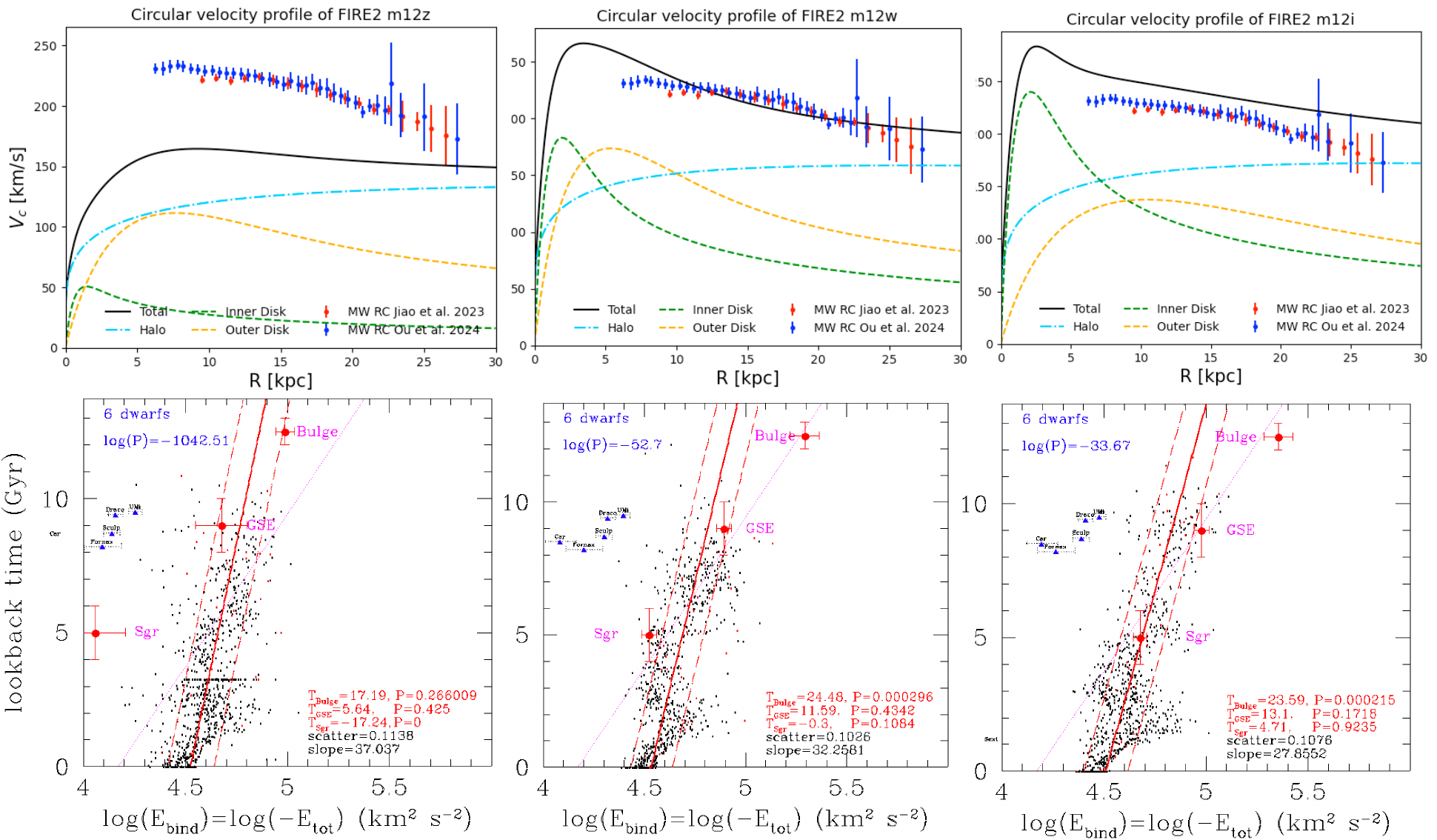}
\includegraphics[width=16cm]{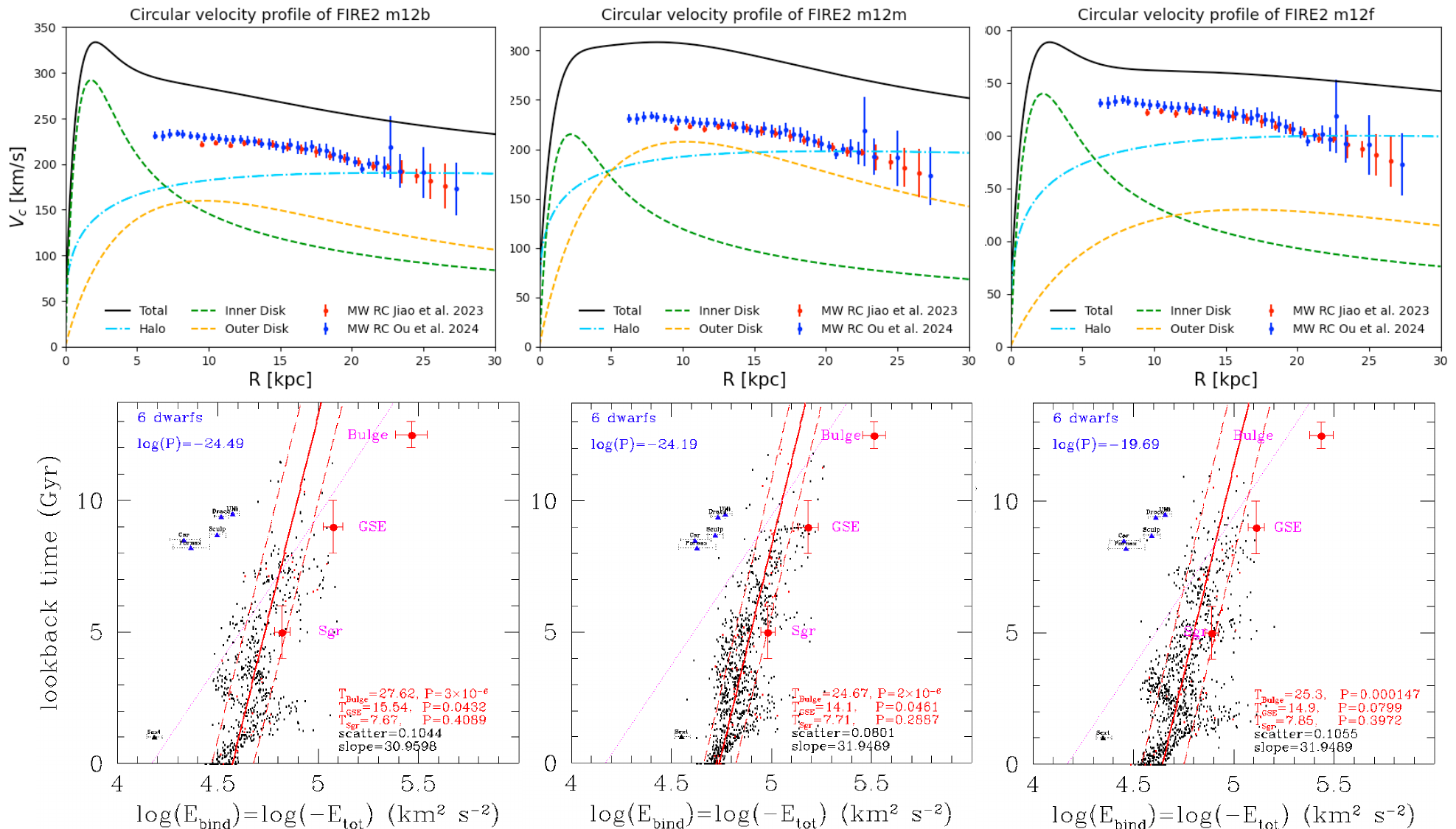}
      \caption{Rotation curves (top panels) and accretion history (bottom panels) from isolated MW-mass galaxies from {\it FIRE}. Same symbols for points and lines than in Figure~\ref{fig:m12c}. For simplicity, here we have only show the six classical dwarfs (blue triangles) and bulge, GSE and Sgr GCs (red crosses).  The top (bottom) panels include simulated galaxies with total mass $M_{\rm 200c}$ smaller (higher) than $10^{12}\,\msun$.
              }
         \label{FigB3}
   \end{figure*}

\begin{figure*}
   \centering
\includegraphics[width=16cm]{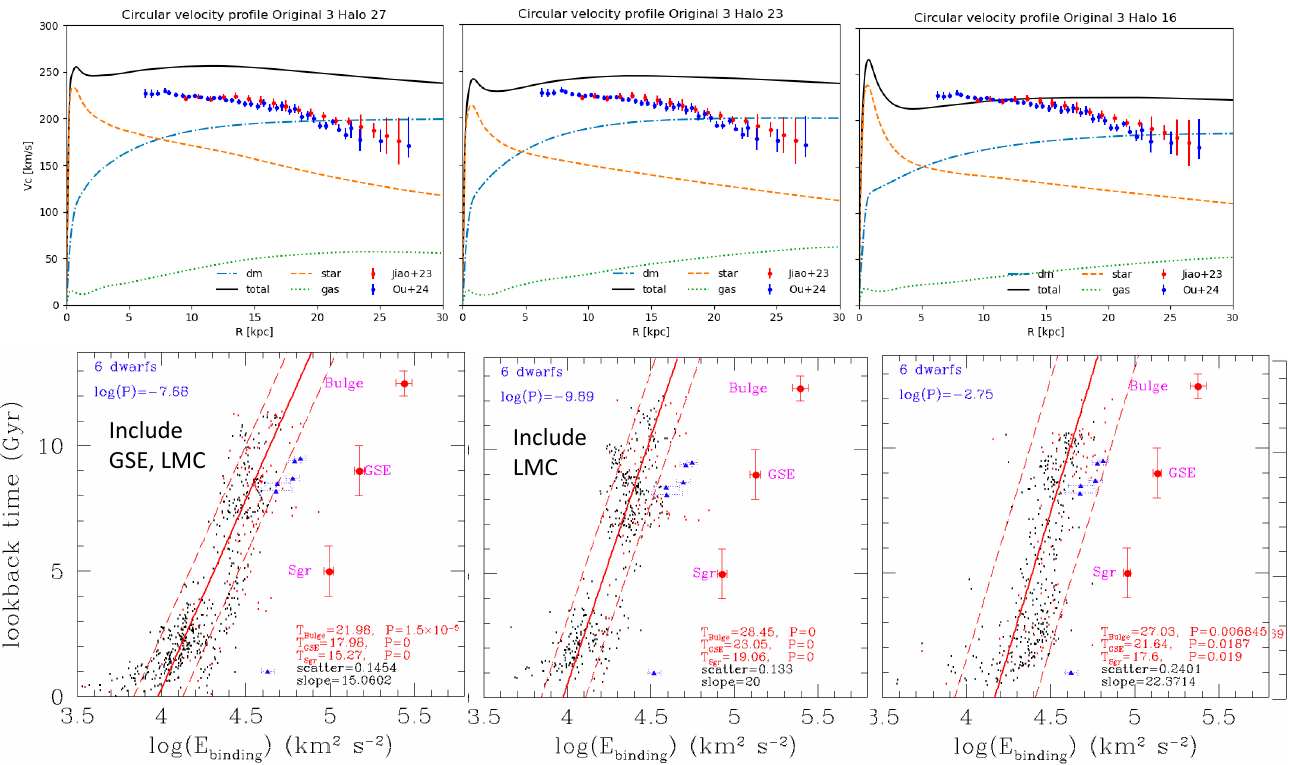}
\includegraphics[width=16cm]{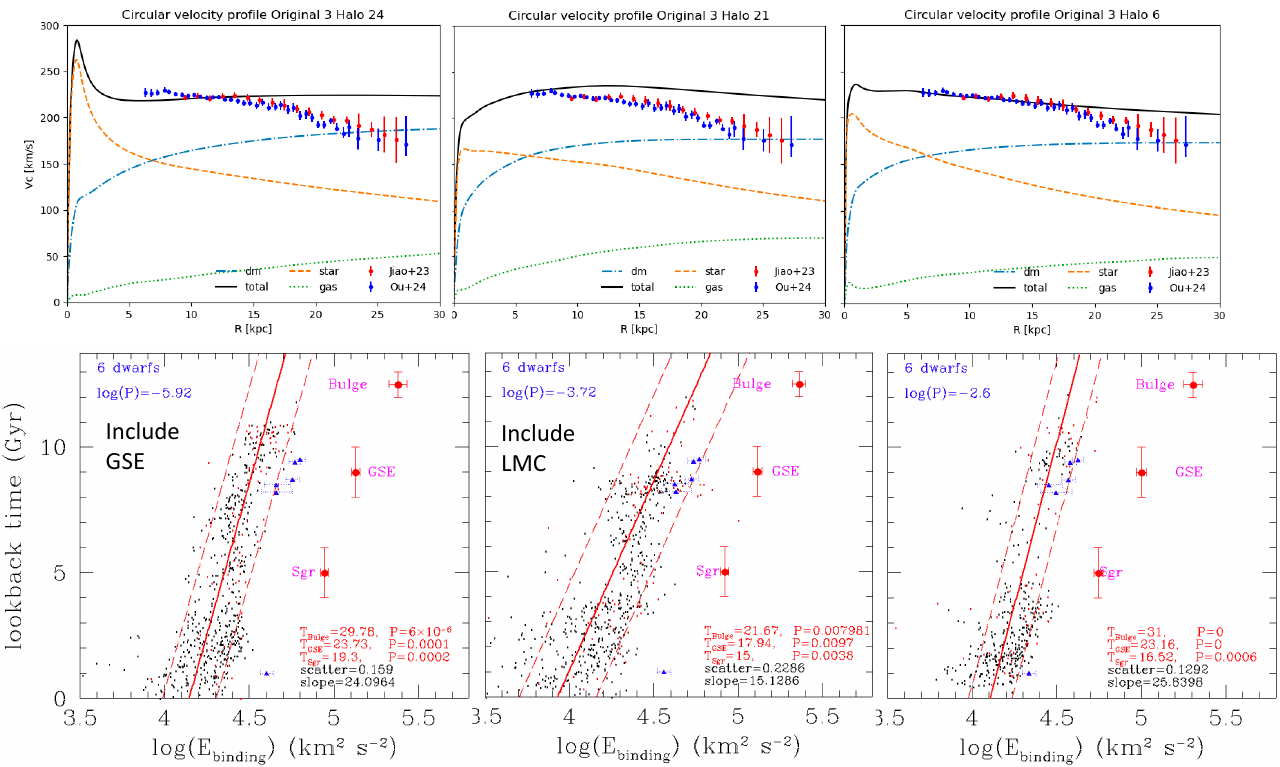}
      \caption{Rotation curves (top) and accretion history (bottom) from high MW-mass galaxies from {\it Auriga}. Same symbols for points and lines than in Figure~\ref{FigB3}. Some {\it Auriga} simulations have been considered to be representative of the infall of GSE or of the LMC, which is indicated in the bottom panel of the corresponding simulations.
              }
         \label{FigB4}
   \end{figure*}

\begin{figure*}
   \centering
\includegraphics[width=\hsize]{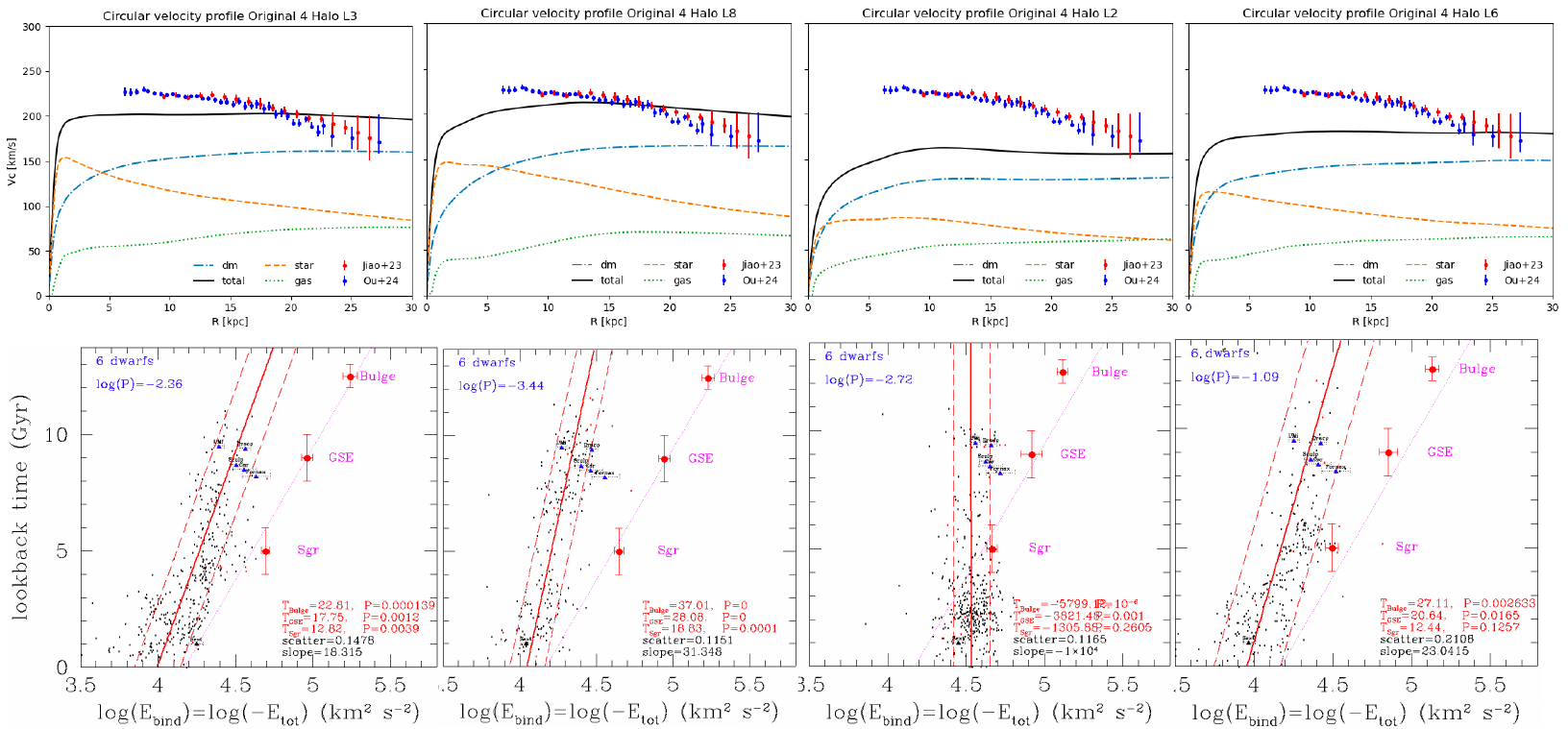}\\
\bigskip
\includegraphics[width=\hsize]{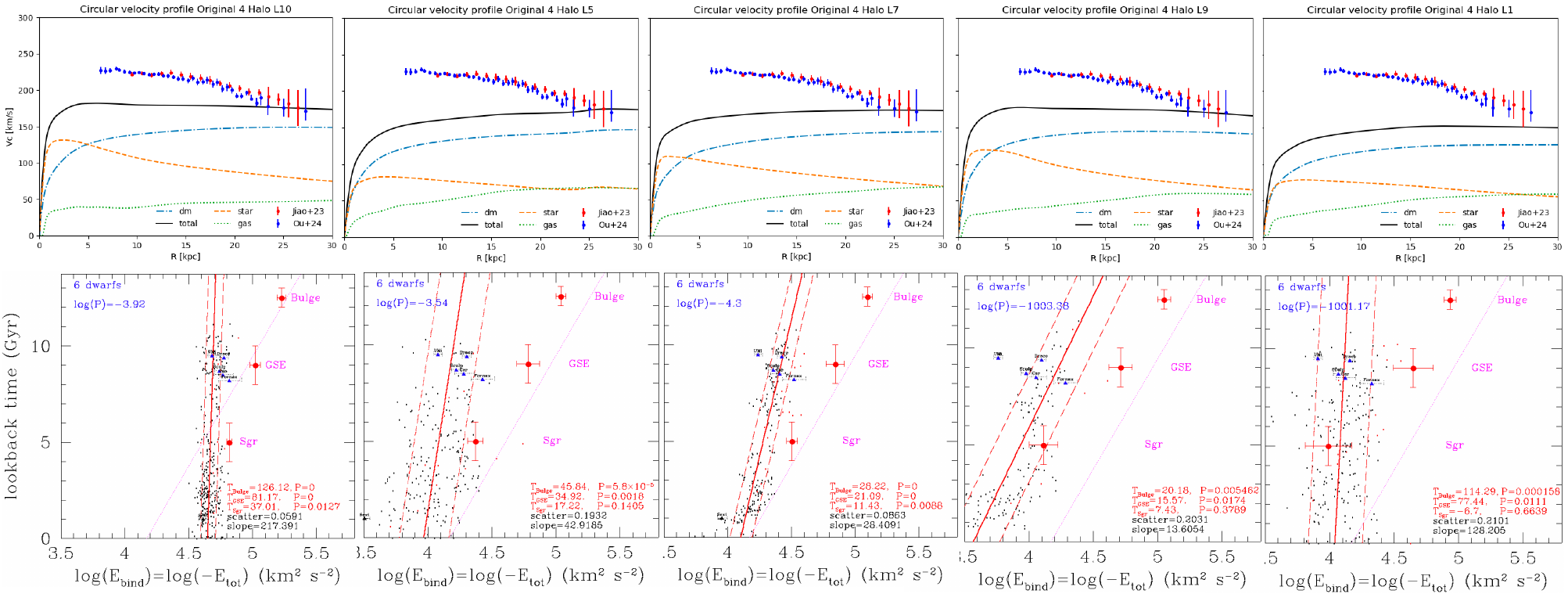}
      \caption{Rotation curves (top) and accretion history (bottom) from low MW-mass galaxies from {\it Auriga}. Same symbols for points and lines than in Figure~\ref{FigB4}. 
              }
         \label{FigB5}
   \end{figure*}

To do this, we have generated the whole orbital history of subhalos and compare their orbits to the evolution of $R_{\rm 200}$. \citet{Rocha2012} used the virial radius $R_{\rm 200m}$ for estimating the infall times as well as it has been done by {\it FIRE}, while {\it Auriga} provides the virial radius $R_{200c}$ at which the mean density is 200 times the critical density. For consistency reasons, we have calculated the virial radius $R_{200\rm m}$ of 15 MW-like simulations from {\it Auriga} at $z=0$ (the last snapshot) with $\Omega_m=0.30966$ \citep[]{Planck2020}. We perform a linear fit on these radii and then roughly convert all $R_{\rm 200c}$ values of all simulations at each redshift to $R_{\rm 200m}$ using this linear relation as presented in Figure~\ref{r200c_r200m}. Finally we assume the infall epoch to be that of the last entry of the subhalo within $R_{\rm 200m}$ as shown in the top panel of Figure~\ref{r200_lookback}. Simulation predictions for both RC and MW accretion history are given in Figure~\ref{FigB3} for {\it FIRE}, and in Figures~\ref{FigB4} (6 high-mass MW with high-resolution) and ~\ref{FigB5} (9 low-mass MW with low-resolution) for {\it Auriga}.\\

During this exercise, we have realized that some subhalos were lacking of information about their orbital positions (e.g. bottom panel in Figure~\ref{r200_lookback}), which render unreliable the infall time estimate. Going one step further, we found this to occur essentially for high-resolution (resolution 3, $5 \times 10^{4}\,\msun$ per DM particle) simulated subhalos with less than 100 particles. Therefore, we have adopted a secure mass limit for subhalos to be $10^{7} \msun$, a value we have also adopted for {\it FIRE} simulated subhalos ($3.5 \times 10^{4}\,\msun$ per DM particle). We have also use the same mass limit for low-resolution (resolution 4, $4 \times 10^{5}\,\msun$ per DM particle) low-mass, {\it Auriga} subhalos because otherwise their numbers become too small to allow an efficient fit of the MW accretion history relation. It  means that the latter relation is less precise for low-mass {\it Auriga} halos L1 to L10 (the last 10 pair of panels in Figure~\ref{FigB5}).

 \end{appendix}

\end{document}